\documentclass{PoS}

\title{Dark Matter Searches with HAWC}

\ShortTitle{Dark Matter Searches with HAWC}

\author{Tolga Yapici\thanks{Corresponding author},$^a$ and \speaker{Andrew J. Smith}$^b$ for the HAWC Collaboration \footnote{Complete list of authors at http://www.hawc-observatory.org/collaboration/icrc2017.php} \\
        \llap{$^a$} University of Rochester \\ E-mail: \email{tyapici@ur.rochester.edu} \\
        \llap{$^b$} University of Maryland, College Park \\ E-mail: \email{asmith8@umd.edu}}

\abstract{The High Altitude Water Cherenkov (HAWC) gamma-ray observatory is a wide field-of-view observatory sensitive to 0.5 TeV - 100 TeV gamma-rays and cosmic-rays in the State of Puebla, Mexico at an altitude of 4100m. The HAWC observatory performed an indirect search for dark matter via GeV-TeV photons resulting from dark matter annihilation and decay considering various sources, including dwarf spheroidal galaxies (dSphs), the M31 galaxy and the Virgo cluster, as well as a combined limit using the dSphs. HAWC has not seen statistically significant excess from these sources. We searched for dark matter annihilation and decay at dark matter masses above 1 TeV. We will present the annihilation cross-section and decay lifetime limits.}

\FullConference{35th International Cosmic Ray Conference --- ICRC217\\
		10--20 July, 2017\\
		Bexco, Busan, Korea}

\usepackage{lineno}
\usepackage{subfigure}

\begin{document}

\section{Introduction}
Despite the fact that dark matter should exist in the universe, it composition is still in question. Among the possible dark matter candidates, Weakly Interacting Massive Particles (WIMPs) are one of the leading hypothetical particle physics candidates for cold dark matter. A WIMP is a dark matter particle that interacts with known standard model particles via a force similar in strength to the weak force \cite{susyDM}.  WIMPs can annihilate into standard model particles and produce photons. WIMPs can also have large dark matter decay lifetimes and would produce gamma rays in similar quantity and energy to the observed neutrinos, from 100 TeV to several PeV~\cite{Kopp:2015bfa,Boucenna:2015tra}. WIMP-like particles which decay may be responsible for the observation of an astrophysical neutrino excess by the IceCube detector~\cite{Aartsen:2013bka,Aartsen:2013jdh,Esmaili:2013gha,Bai:2013nga}.

Among the places in the Universe to look for signatures of dark matter, dwarf spheroidal galaxies (dSphs) are some of the best candidates for a dark matter search due to their high dark matter content and low luminous material. The dwarf spheroidal galaxies considered in this analysis are companion galaxies of the Milky Way, in what is known as our Local Group. They are very low luminosity galaxies, with low diffuse Galactic gamma-ray foregrounds and little to no astrophysical gamma-ray production~\cite{2015arXiv151000389B}. A total of 15 dSphs are considered in this analysis: Bootes I, Canes Venatici I, Canes Venatici II, Coma Berenices, Draco, Hercules, Leo I, Leo II, Leo IV, Segue 1, Sextans, Ursa Major I, Ursa Major II, Ursa Minor and TriangulumII. These dSphs were chosen for their favored declination angle for the HAWC observatory and well studied dark matter content with respect to other dSphs. 

The best DM decay lifetime limits are expected to come from the most massive objects in the universe such as galaxies and galaxy clusters, thus we studied M31 galaxy and the Virgo Cluster. Even though these sources are extended sources ($\sim$3-6$^\circ$), in this analysis, we treated them as point sources. More detailed analyses incorporating their DM morphology are in progress and they should provide better limits than presented here.

\section{HAWC Observatory}

The High Altitude Water Cherenkov (HAWC) observatory detects high-energy gamma-ray and is located at Sierra Negra, Mexico. The site is 4100 m above sea level, at latitude 18$^\circ$59.7' N and longitude 97$^\circ$18.6' W. HAWC is a survey instrument that is sensitive to gamma rays of 500 GeV to a few hundred TeV \cite{2017arXiv170101778A} energies. HAWC consists of 300 water Cherenkov detectors (WCDs) covering 22000 m$^2$ area. Each detector contains four photo-multiplier tubes \cite{2017arXiv170101778A} and the trigger rate of HAWC is approximately 25kHz. HAWC has a duty cycle >95\% and a wide, unbiased field of view of $\sim$2 sr. 

It has been operating with a partial detector since August 2013 and has been operating with the full detector since March 2015. Here we present results from 507 days of its operations with the full detector.

\section{Dark Matter Gamma-ray Flux}
\subsection{Gamma-ray Flux from Dark Matter Annihilation and Decay}

Expected gamma-ray fluxes from dark matter annihilation were calculated using both the astrophysical properties of the potential dark matter source and the particle properties of the initial and final-state particles for different dark matter masses and for different annihilation channels. The differential gamma-ray flux integrated over solid angle of the source is given by:

\begin{equation} \label{Flux}
\frac{\mathrm{d}F}{\mathrm{d}E}_{annihilation} = \frac{\langle\sigma_{A}v\rangle}{8\pi M_{\chi}^{2}}\frac{\mathrm{d}N_{\gamma}}{\mathrm{d}E}J
\end{equation}
where $\langle\sigma_{A}v\rangle$ is the velocity-weighted dark matter annihilation cross-section, $dN_{\gamma}/dE$ is the gamma-ray spectrum per dark matter annihilation, and $M_{\chi}$ is the dark matter particle mass. $J$-factor ($J$) is defined as the dark mass density ($\rho$) squared and integrated along the line of sight distance $x$ and over the solid angle of the observation region:
\begin{equation} \label{Jdeltaomega}
J= \int_{\rm source} \mathrm{d}\Omega \int \mathrm{d}x \rho^{2} (r(\theta,x)) 
\end{equation}
where the distance from the earth to a point within the source is given by
\begin{equation} \label{rgal}
r(\theta,x) = \sqrt{R^{2} - 2xR\cos(\theta) + x^{2}}\enspace,
\end{equation}
$R$ is the distance to the center of the source, and $\theta$ is the angle between the center of the source and the line of sight. 

The gamma-ray flux from dark matter decay is similar to the dark matter annihilation gamma-ray flux as described above in Equation \ref{Flux}:

\begin{equation} \label{Flux_decay}
\frac{\mathrm{d}F}{\mathrm{d}E}_{decay} = \frac{1}{4\pi \tau M_{\chi}}\frac{\mathrm{d}N_{\gamma}}{\mathrm{d}E}D\enspace.
\end{equation}

Decay process involves only one particle, thus, the gamma-ray flux from dark matter decay depends on a single power of the dark matter density $\rho$ instead of the square:

\begin{equation}
D=\int_{\rm source} d\Omega \int dx \rho (r_{gal}(\theta,x))\enspace.
\end{equation}


\subsection{Dark Matter Density Distributions}\label{DMdenssec}

Density profiles describe how the density ($\rho$) of a spherical system varies with distance ($r$) from its center. We used the Navarro-Frenk-White (NFW) model for the dark matter density profiles. The NFW density profile is given by

\begin{equation} \label{NFW}
\rho_{\rm NFW} (r)= \frac{\rho_{s}}{(r/r_{s})^\gamma(1+(r/r_{s})^\alpha)^{(\beta-\alpha)/\gamma}}
\end{equation}
where $\rho_{s}$ is the scale density, $r_{s}$ is the scale radius of the galaxy, $\gamma$ is the slope for $r<<r_s$, $\beta$ is the slope for $r>>r_s$ and $\alpha$ is the transition parameter from inner slope to outer slope. NFW profiles parameters from~\cite{Geringer-Sameth2015} are used for all sources listed, except for Triangulum II, for dSphs. The J- and D-factor for 14 dSph sources are calculated using the {\sc CLUMPY} software ~\cite{Bonnivard2016} for different realizations of the tabulated values and their respective uncertainties for an angular window of $\theta_{max}$ from \cite{Geringer-Sameth2015}.  A similar procedure was applied for M31 \cite{Tamm2012} and the Virgo Cluster \cite{Sanchez-Conde2014}. The mean values of the J- and D-factors are tabulated in Table \ref{table:sourceparameters}. For Triangulum II, we use the $J$ and $D$ factors from~\cite{Hayashi:2016kcy}.

\begin{table}
  \centering
  \begin{tabular}{ccccc}
  	\hline
    Source  & RA & Dec & log$_{10}$ (J/$\mathrm{GeV}^2 \mathrm{cm}^{-5} \mathrm{sr}$) & log$_{10}$ (D/$\mathrm{GeV} \mathrm{cm}^{-5} \mathrm{sr}$) \\ \hline\hline
    Bootes1         & 210.05 & 14.49 & 18.47 & 18.45 \\ 
    CanesVenaticiI  & 202.04 & 33.57 & 17.62 & 17.55 \\
    CanesVenaticiII & 194.29 & 34.32 & 17.95 & 17.68 \\ 
    ComaB           & 186.74 & 23.90 & 19.32 & 18.71 \\ 
    Draco           & 260.05 & 57.07 & 19.37 & 19.15 \\
    Hercules        & 247.72 & 12.75 & 16.93 & 16.87 \\
    LeoI            & 152.11 & 12.29 & 17.57 & 18.04 \\
    LeoII           & 168.34 & 22.13 & 18.11 & 17.33 \\
    LeoIV           & 173.21 & -0.53 & 16.37 & 16.50 \\
    Segue1          & 151.75 & 16.06 & 19.66 & 18.64 \\
    Sextans         & 153.28 & -1.59 & 17.96 & 18.59 \\
    TriangulumII    & 33.32  & 36.18 & 20.44 & 18.42 \\
    UrsaMajorI      & 158.72 & 51.94 & 18.66 & 18.11 \\
    UrsaMajorII     & 132.77 & 63.11 & 19.67 & 19.05 \\
    UrsaMinor       & 227.24 & 67.24 & 19.24 & 17.92 \\\hline
    M31             & 10.68  & 41.27 & 20.86 & 19.10 \\
    Virgo Cluster   & 186.75 & 12.38 & 19.50 & 19.44 \\\hline
\end{tabular}
 \caption{Astrophysical parameters, J-factors and D-factors, for the fifteen dwarf spheroidal galaxies within the HAWC field-of-view, M31 and Virgo Cluster. The source, right ascension ($RA$), declination ($Dec$), the dark matter $J$-factor and $D$-factor are listed above. NFW profiles parameters from~\cite{Geringer-Sameth2015} are used for all dSphs listed, except for Triangulum II. For Triangulum II, we use the $J$ and $D$ factors from~\cite{Hayashi:2016kcy}. For M31 and the Virgo Cluster, the profile parameters from \cite{Tamm2012} and \cite{Sanchez-Conde2014} are used, respectively. J-factors and D-factors are calculated for integration angle of $\theta_{max}$ for respective sources.}
  \label{table:sourceparameters}
\end{table}
\subsection{Systematics}

Systematic uncertainties arise from a number of sources within the detector. These effects were carefully investigated in \cite{2017arXiv170101778A}. The overall systematic uncertainty on the HAWC data set is on the order of $\pm50\%$ on the observed flux. The uncertainties on the expected dark matter annihilation and decay limits were calculated to account for these systematics uncertainties.

There are additional systematic uncertainties on the expected dark matter flux due to the integration angle of J- and D-factor. HAWC has an angular resolution between 1$^\circ$ and 0.2$^\circ$ for near zenith angles. In the case of better angular resolutions, the integration angle gets smaller which makes J- and D-factors smaller. Similarly, for worse angular resolutions cases, the integration angle gets greater that makes the J- and D- factors larger values. However, there is constraint on the dark matter distribution which limits the dark matter content of a source at angles larger than $\theta_{max}$. We impose this physically motivated constraint on the J- and D- factor uncertainties. For combined limit uncertainties, we used the uncertainties corresponding to Segue1 (42\% for annihilation cross-section limits and 38\% for decay lifetime limits) since it is one of the dominant sources. 

\section{Limits on the Dark Matter Annihilation Cross Section and Decay Lifetime with HAWC data}

We analyzed the individual and combined limits from 15 dwarf spheroidal galaxies within the HAWC field of view, M31 galaxy and the Virgo Cluster for the HAWC 507 days data. Considering the angular resolution of HAWC observatory ($\sim$0.5 degrees) \cite{2017arXiv170101778A}, the limits were calculated assuming that the dSphs are point sources. No statistically significant gamma-ray flux has been found for a range of dark matter masses, 1~TeV -- 100~TeV, and five dark matter annihilation channels. So, we calculated $95\%$ confidence level limits on the annihilation cross-section and decay lifetime, the source significance is used to determine the exclusion curves on the dark matter annihilation cross-section $\langle \sigma_{A}v\rangle$ and decay lifetime $\tau$, for the individual sources. A joint likelihood analysis was also completed by combining the statistics for all 15 dSphs in order to increase the sensitivity of the analysis. In this paper, we only present results for $\tau\bar{\tau}$ channel. The rest of the results can be found in \cite{2017arXiv170601277A} with the explanation of the limit calculations. For other possible studies, flux upper limits were also presented in the respective paper.

\begin{figure}[t!]
\subfigure[]{
  \includegraphics[width=0.499\textwidth]{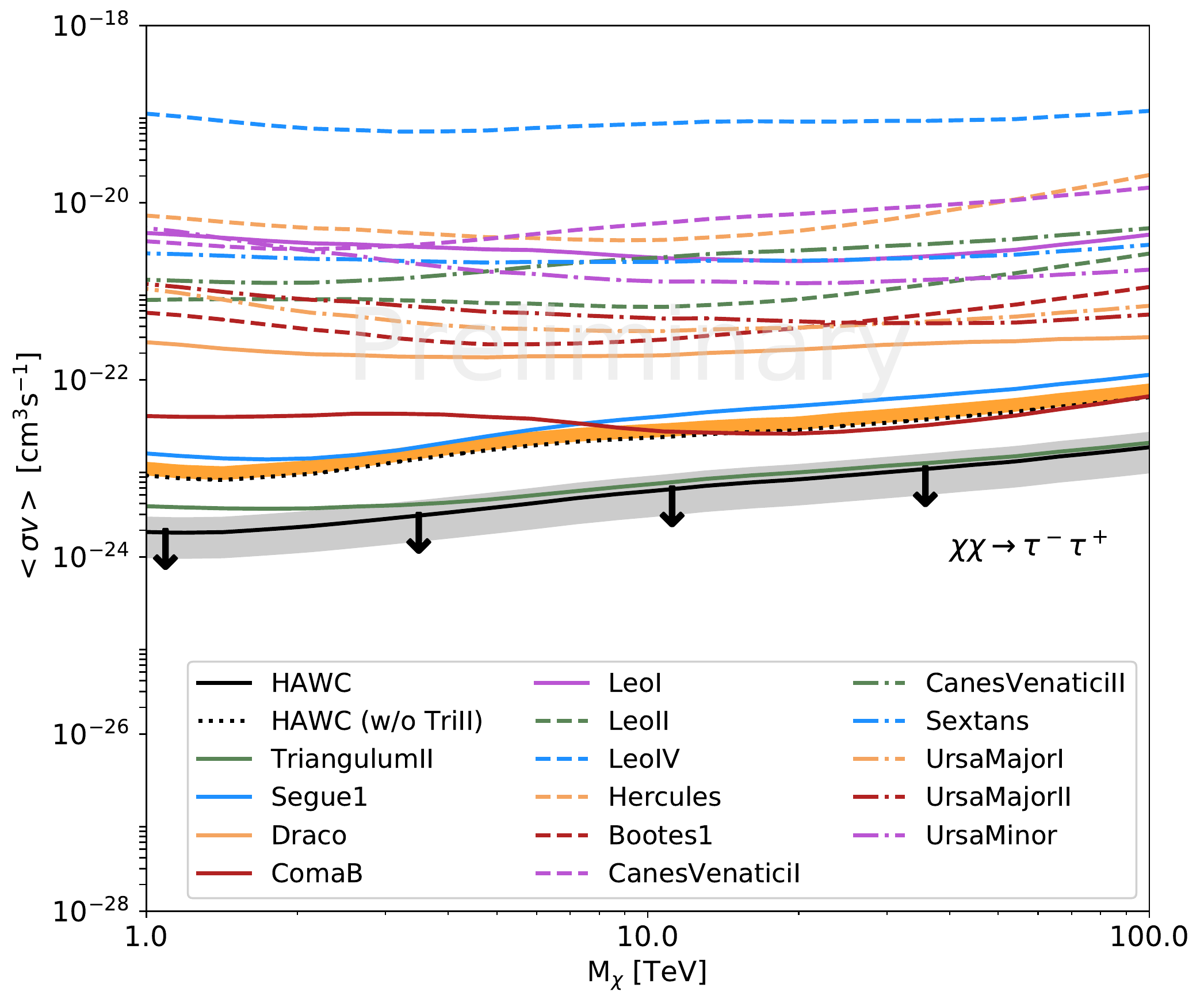}
}
\subfigure[]{
\includegraphics[width=.499\textwidth]{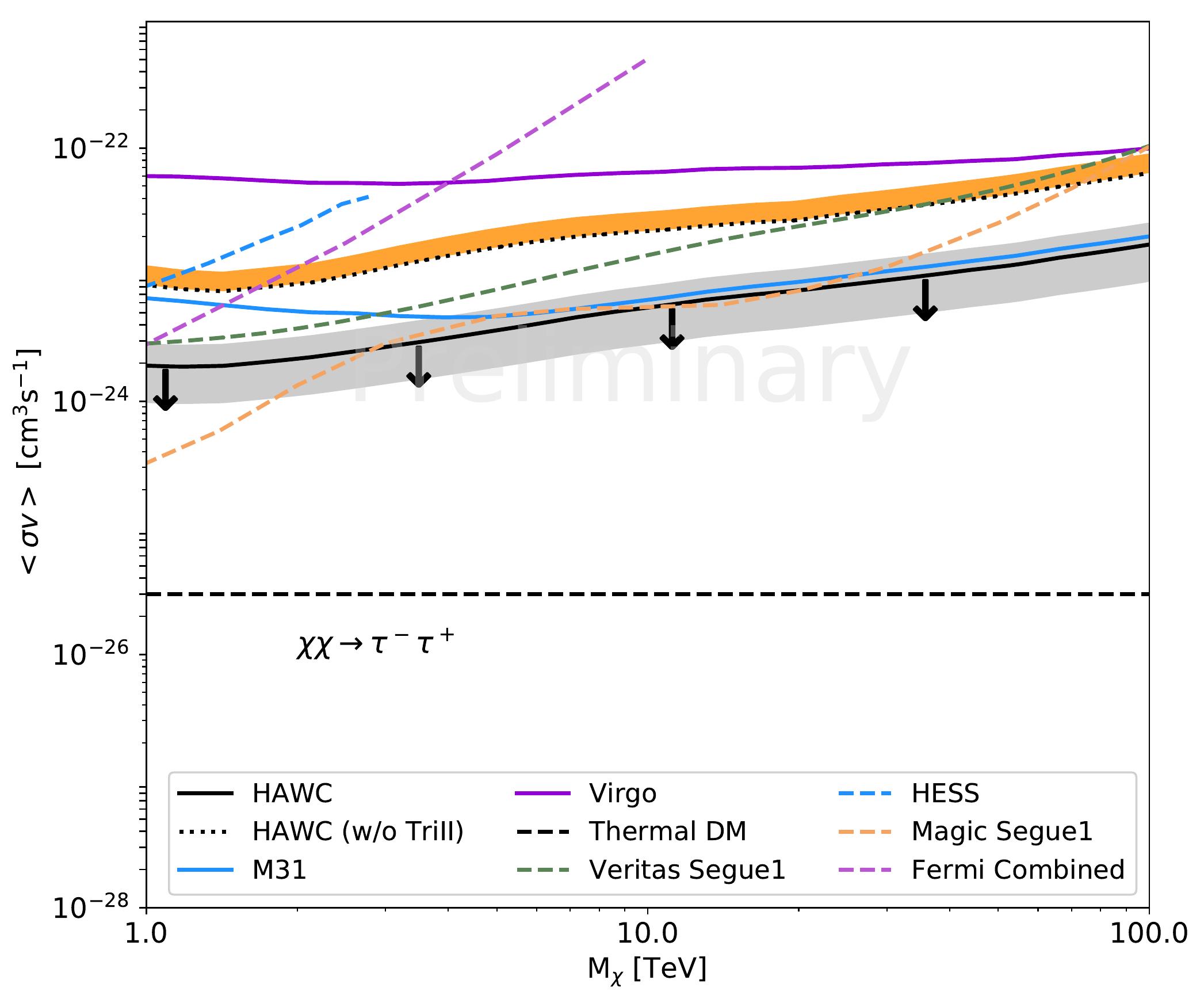}
}
\caption{95$\%$ confidence level upper limits on the dark matter annihilation cross-section (a) for 15 dwarf spheroidal galaxies within the HAWC field of view for the $\tau^{+}\tau^{-}$ annihilation channels and (b) its comparison to other experimental results, Fermi-LAT combined dSph limits \cite{Fermi2014}, Veritas Segue 1 limits \cite{2017PhRvD..95h2001A}, HESS combined dSph limits \cite{HESS2014} and MAGIC Segue 1 limits \cite{MAGIC2016} are shown along with the HAWC M31 and Virgo Cluster limits. The gray band shows the systematic uncertainty on the combined limits due to HAWC systematics and dark orange band shows the systematic uncertainty due to J-factor uncertainty.
\label{15dwarffigs_annihilation}}
\end{figure}

\begin{figure}[b!]
\subfigure[]{
  \includegraphics[width=0.499\textwidth]{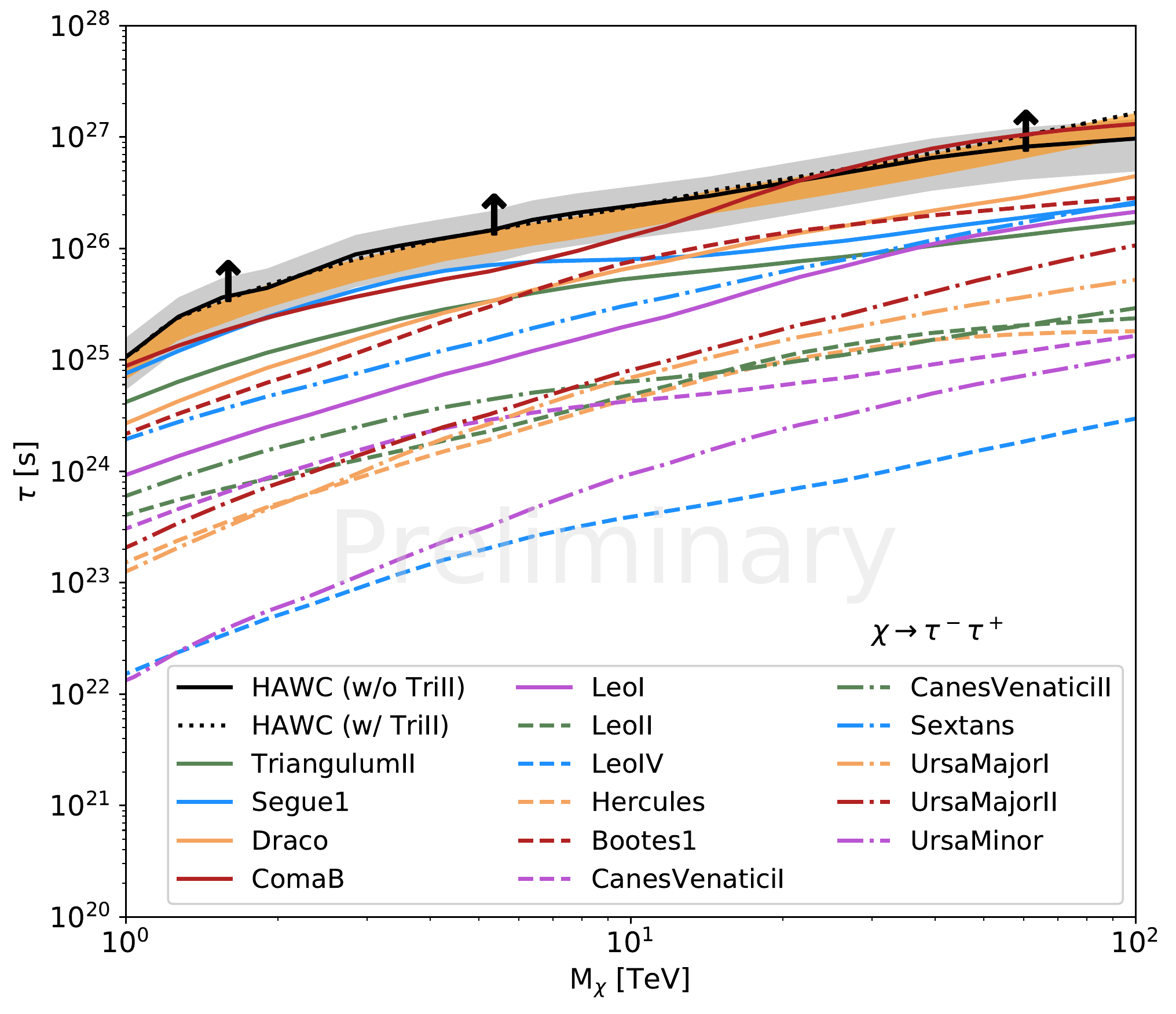}
}
\subfigure[]{
	\includegraphics[width=0.499\textwidth]{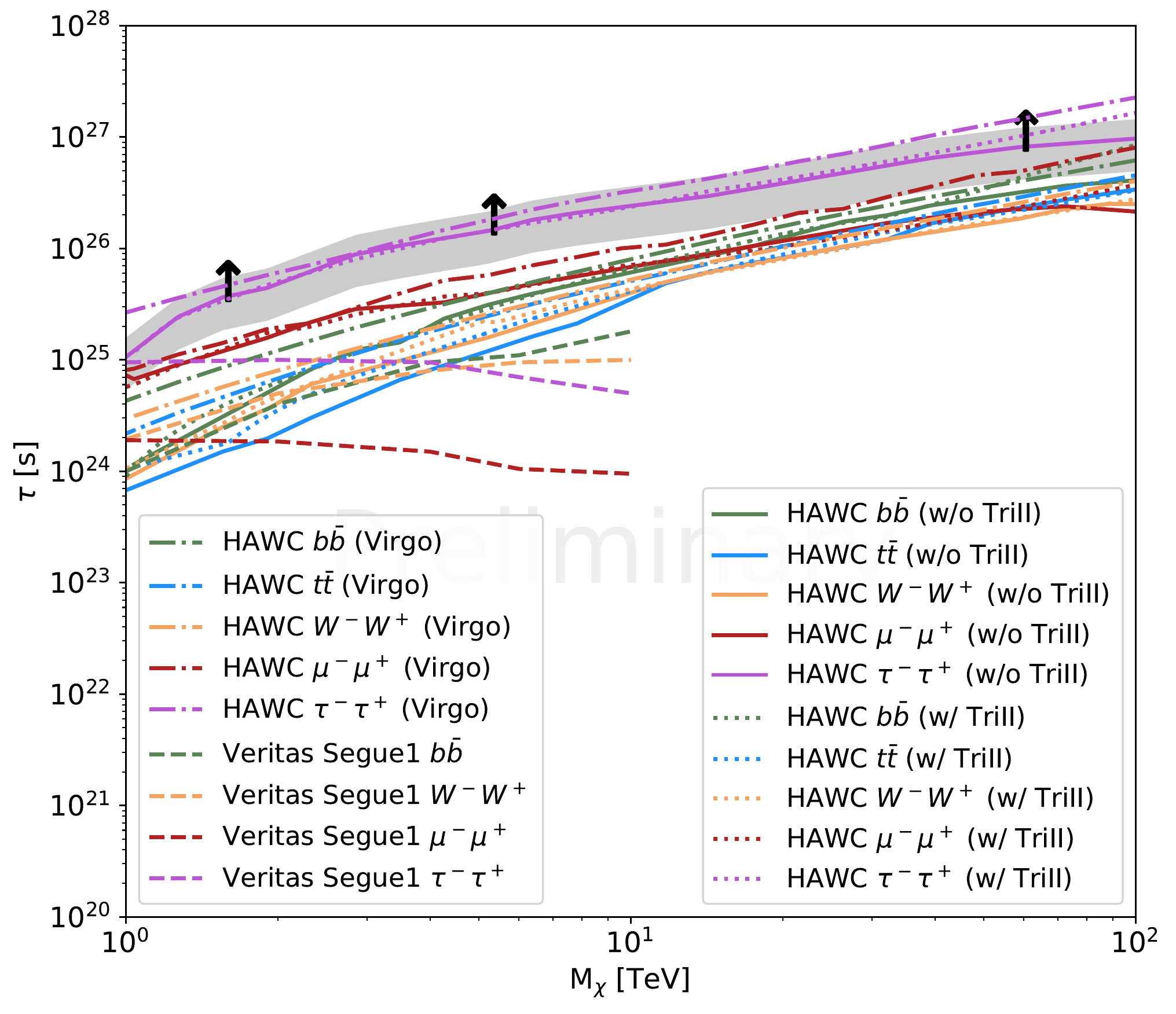}
}  
	\caption{95$\%$ confidence level lower limits on the dark matter decay lifetime (a) for 15 dwarf spheroidal galaxies within the HAWC field of view and (b) for combined, M31 and Virgo Cluster limits compared with Veritas Segue 1 limits \cite{veritas2012} for the $\tau^{+}\tau^{-}$. The gray band shows the systematic uncertainty on the combined limits due to HAWC systematics and dark orange band shows the systematic uncertainty due to D-factor uncertainty. 
    \label{fig:15dwarffigs_decay}}
\end{figure}

Triangulum II has a particularly large $J$-factor and transits near the zenith for HAWC. However, it was discovered recently~\cite{2015ApJ...802L..18L} and it still has large uncertainties in its DM profile. Because of this, we show the joint dwarf limit both with and without Triangulum II in Figures~\ref{15dwarffigs_annihilation} and~\ref{fig:15dwarffigs_decay}.


For the $b\bar{b}$ channel, Fermi-LAT limit is the most contraining up to $\sim$4 TeV, the followed by MAGIC Segue 1 limit up to $\sim$10 TeV. Beyond $\sim$10 TeV, the HAWC combined dSph limit is the most stringent limit for this channel. The HAWC combined limits with Triangulum II are strongest for the $W^+W^-$ channel for ${M_\chi\gtrsim\mbox{30 {\rm TeV}}}$. This result is consistent within uncertainties with Veritas Segue 1 limit. For $\mu^+\mu^-$ and $\tau^+\tau^-$ channels, the HAWC combined dSph limits are the strongest above a few TeV. M31 limits are also comparable with the combined limits with Triangulum II whereas the Virgo Cluster is not as sensitive as M31 galaxy (Figure \ref{15dwarffigs_annihilation})

Dark matter models for thermal relic and Sommerfeld enhanced cross-sections are compared. For the Sommerfeld enhancement, a weak-scale coupling of 1/35 and a very conservative dark matter velocity of 300\,km/s was assumed. Only $W^+W^-$ annihilation channel is considered for the Sommerfeld enhancement since this channel is assured to have dark matter coupled to gauge bosons \cite{sommerfeld}. At resonances, HAWC limit rules out a dark matter with mass of $\sim$4 TeV, and HAWC limit approaches to corresponding Sommerfeld-enhanced models by 1 order of magnitude for a dark matter with mass of $\sim$20 TeV. Slower dark matter velocity enhances the amplitude of resonances, thus making HAWC results closer to Sommerfeld-enhanced thermal relic.

Figure \ref{fig:15dwarffigs_decay} shows 15 individual dSph, the combined, M31 and the Virgo cluster limits. Like the dark matter annihilation results, Segue 1, Coma Berenices, and Triangulum II are dominant, though for decays, Bootes I and Draco also contribute to the combined limits. This is due to the fact that dark matter decay is related to $\int\rho$ (total dark matter mass) compared to $\int\rho^2$ at the source of annihilation or decay. The strongest lifetime lower limit is obtained with the $\tau^+\tau^-$ channel. The Virgo Cluster results are within 2--3 factors of the combined dSph limits. Despite not providing good limits for annihilation, the good decay lifetime limits for the Virgo Cluster is due to the total DM mass in the cluster. 

\section{Summary}
We present 95\% CL limits on the annihilation cross section and the decay lifetime for 15 dwarf spheroidal galaxies within the HAWC field-of-view, M31 Galaxy and the Virgo Cluster. A combined limit is also shown from a stacked analysis of all dwarf spheroidal galaxies. These are the limits also shown in \cite{2017arXiv170601277A}.

The HAWC collaboration is improving its analysis tools for enhancing energy and angular resolution. In addition, with more data collected, HAWC is expected to be more sensitive at lower dark matter masses.

\section*{Acknowledgments}
\footnotesize{
We acknowledge the support from: the US National Science Foundation (NSF);
the US Department of Energy Office of High-Energy Physics;
the Laboratory Directed Research and Development (LDRD) program of
Los Alamos National Laboratory; Consejo Nacional de Ciencia y Tecnolog\'{\i}a (CONACyT),
Mexico (grants 260378, 232656, 55155, 105666, 122331, 132197, 167281, 167733);
Red de F\'{\i}sica de Altas Energ\'{\i}as, Mexico;
DGAPA-UNAM (grants IG100414-3, IN108713,  IN121309, IN115409, IN111315);
VIEP-BUAP (grant 161-EXC-2011);
the University of Wisconsin Alumni Research Foundation;
the Institute of Geophysics, Planetary Physics, and Signatures at Los Alamos National Laboratory;
the Luc Binette Foundation UNAM Postdoctoral Fellowship program.
}

\bibliographystyle{JHEP}
\bibliography{references}

\end{document}